\documentclass[%
reprint,
showpacs,
showkeys,
preprintnumbers,
nofootinbib,
nobibnotes,
amsmath,amssymb,aps,prl]{revtex4-1}
\usepackage{graphics}
\usepackage{epsfig}
\usepackage{dcolumn}
\usepackage{color}
\usepackage{bm}
\usepackage{subfigure}
\usepackage{hyperref}
\usepackage[mathlines]{lineno}

\begin{document}


\title{Quantum Hall Edge States in Topological Insulator Nanoribbons} 

\author{A.~Pertsova}
\altaffiliation[Now at: ]{Nordita, KTH Royal Institute of Technology and Stockholm University, Roslagstullsbacken 23, SE-106 91 Stockholm, Sweden}
\email{anna.pertsova@su.se}
\author{C. M. Canali}
\affiliation{Department of Physics and Electrical Engineering, Linn{\ae}us University, 391 82 Kalmar, Sweden}
\author{A.~H.~MacDonald}
\affiliation{Department of Physics, University of Texas at Austin, TX 78712, USA}


\begin{abstract}
We present a microscopic theory of the chiral one-dimensional 
electron gas system localized on the sidewalls of  
magnetically-doped Bi$_2$Se$_3$-family topological insulator nanoribbons in
the quantum anomalous Hall effect (QAHE) regime.
Our theory is based on a simple continuum model of sidewall states whose 
parameters are extracted from detailed ribbon and film geometry 
tight-binding model calculations.  
In contrast to the familiar case of the quantum Hall effect in 
semiconductor quantum wells, the number of microscopic 
chiral channels depends simply and systematically on the ribbon thickness and on the position of 
the Fermi level within the surface state gap.  We use our theory to interpret recent transport experiments 
that exhibit non-zero longitudinal resistance in samples with accurately quantized Hall conductances.
\end{abstract}


\maketitle

\noindent
\textit{Introduction}---The quantum Hall effect~\cite{KlitzingPhysRevLett.45.494} is a transport anomaly 
that occurs when~\cite{MacDonald1995} a two-dimensional electron system 
has a charge gap, {\it i.e.} a jump in chemical potential, at a density that depends on magnetic field.
It is characterized by the absence of longitudinal resistance and quantized Hall resistance.  Both properties
can be understood in terms of the chiral one-dimensional 
electron systems~\cite{Wen1991} (C1DESs) always present at quantum Hall sample 
edges.  Although the rate at which their non-zero equilibrium currents change as chemical potential is varied
is fixed by the magnetic field dependence of the gap density, other properties of C1DESs are dependent on microscopic details.  
In the case of GaAlAs two-dimensional electron gas systems, for example, it has in fact been difficult to achieve 
a fully satisfactory understanding of chiral edge state properties because of electrostatic imperatives 
that force edge reconstructions~\cite{Chamon1994} and cause the number of microscopic edge channels to 
proliferate~\cite{Chklovskii1992}.  Accurate quantization of the Hall conductance then requires~\cite{MacDonald1995} 
only that local equilibrium be established at decoupled edges 
of the sample.   

In this Letter we address the properties of the chiral one-dimensional electron gas system associated 
with the quantum anomalous Hall effect (QAHE)~\cite{HaldanePhysRevLett.61.2015,LiuPhysRevLett.101.146802,Yu02072010,Chang12042013,WangPhysRevLett.111.086803,QiPhysRevB.74.085308,QiPhysRevB.78.195424,LiuPhysRevLett.101.146802,NomuraPhysRevLett.106.166802,Yu02072010,WangPhysRevLett.111.086803,He31122013,KouSSC2015,WengQAHE2015} 
in magnetically doped topological insulator~\cite{Hasan,XLQi} thin films.
The appearance of a quantum Hall effect in these systems is a direct 
consequence of spontaneously broken time-reversal symmetry, which is 
also manifested by a suite of unusual 
magnetic~\cite{TserkovnyakPhysRevLett.108.187201,ZhuPhysRevLett.106.097201,Chen06082010,Checkelsky2012,Xu2012,GrauerPhysRevB.92.201304}, 
and optical~\cite{TsePhysRevLett.105.057401,MalshukovPhysRevB.88.245122,LasiaPhysRevB.90.075417} 
properties.  We show that the chiral one-dimensional electron system associated with this quantum 
Hall effect is localized on the thin film side walls and that, in contrast to the case of GaAs quantum wells, 
its microscopic properties depend rather simply on film thickness, the size of the surface state gap induced by 
broken time-reversal symmetry, and also on the facet dependence of surface-state Dirac-cone velocities.  
Using our theory, we argue that in the absence of disorder thin films with the characteristics
of samples in which the QAHE has so far been studied support a 
C1DES with a single chiral channel.  It follows that 
the presence of a non-zero longitudinal resistance in most 
experiments~\cite{Chang12042013,CheckelskyNatPhys2014,KouPhysRevLett.113.137201,
BestwickPhys.Rev.Lett.114.187201,ChangNatPhys2015,ChangPhysRevLett.115.057206,GrauerPhysRevB.92.201304}
cannot be attributed, as is common, to the absence of local equilibrium at a multi-channel edge.  

\begin{figure}[ht!]
\centering
\includegraphics[width=0.98\linewidth,clip=true]{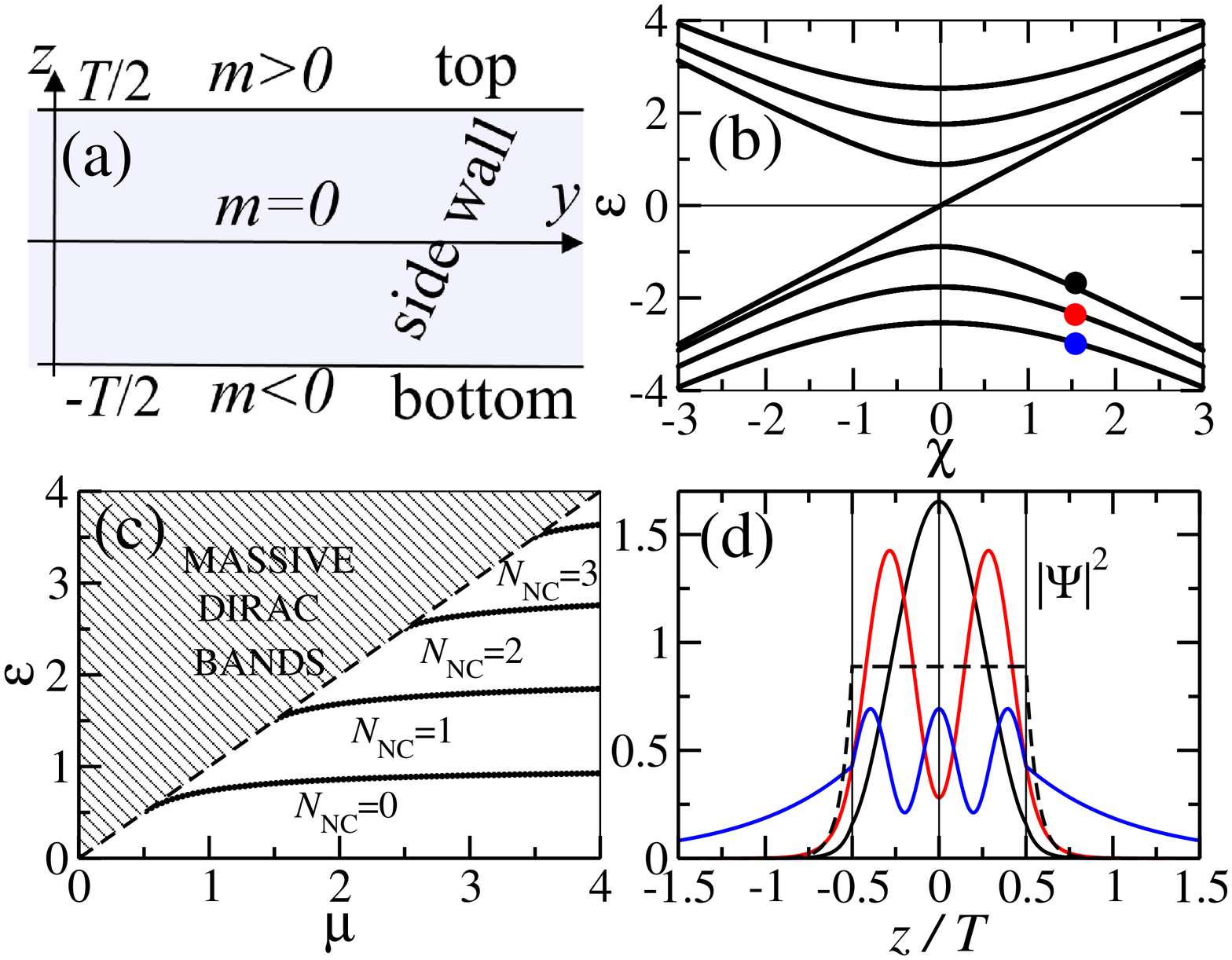}
\caption{(Color online) (a) 
Schematic illustration of a simplified Dirac model for states localized on a $y$-$z$ plane sidewall. 
The mass term $m(z)$ in the Dirac Hamiltonian has opposite 
signs for top ($z>T/2$) and bottom ($z<-T/2$) surfaces and vanishes on
the sidewall surface ($-T/2<z<T/2$). 
(b) Quasi one-dimensional energy bands calculated
at $\mu=8/\pi$ with energies in $\hbar \pi v_{\mathrm{D}z}/ T$ units.
(c) Positive non-chiral band energies at $\chi=0$ as a function of the 
dimensionless gap $\mu$.  The non-chiral bands are particle-hole symmetric.
For $\epsilon > \mu$ all states are extended across
the top and bottom surfaces.    
In the limit $\mu \rightarrow \infty$ the dimensionless non-chiral sidewall
state energies approach integers.  
For $\epsilon < \mu$, the $\chi=0$ band energy lines separate regions labelled by 
the number of non-chiral channels $N_{\rm NC}$ that are present.   
(d) Wavefunctions of the 
chiral state (black dashed line) and the first three negative energy non-chiral states
at positive (dimensionless) momentum $\chi=5/\pi$.  The eigenvalues associated with the
non-chiral wavefunctions are marked by correspondingly colored dots in (b). 
} 
\label{fig4}
\end{figure}

\noindent
\textit{Sidewall State Toy Model}---
A qualitative understanding of C1DES properties 
can be obtained from the simplest possible 2D sidewall model.
(See Refs.~\onlinecite{LouAPL2011,ChenAPL2015,DebJPCM20014,0953-8984-24-1-015004,
BardasonPhysRevLett.105.156803, RosenbergPhysRevB.82.041104,ZhangPhysRevLett.105.206601,
ChangPhysRevLett.106.206802,BreyPhysRevB.89.085305,Acero2015} for 
related continuum model analysis.)
We assume that the sidewall is infinite in extent in the $\hat{y}$ (horizontal) direction,
that it has thickness $T$ in the $\hat{z}$ (vertical) direction [See Fig.~\ref{fig4}(a)], 
and that it is described by an anisotropic Dirac Hamiltonian with a mass term: 
$\hat{H}=i\hbar (- v_{\mathrm{D}z} \sigma_y\partial_z+ v_{\mathrm{D}y} \sigma_x\partial_y)+m(z)\sigma_z$.
Here $\mathbf{\sigma}=\lbrace \sigma_x,\sigma_y,\sigma_z \rbrace$ is a Pauli-matrix vector
that acts on spin, $m(z)$ captures the influence of exchange interactions between
the top, bottom, and sidewall surface quasiparticles and the $\hat{z}$ direction bulk magnetization, 
$v_{\mathrm{D}z}$ is the vertical Dirac velocity, and $v_{\mathrm{D}y}$ the horizontal 
Dirac velocity.  The mass is zero on the side wall 
where the exchange interaction can be absorbed by a gauge change,  
($m(z)=0$ for $-T/2<z<T/2$) and has a different sign 
on the top  and bottom  surfaces; $m(z)=m_0>0$ for $z>T/2$ 
and $m(z)=-m_0<0$ for $z<-T/2$, where $m_0$ is a constant. 
We can find the eigenvalues and eigenvectors of this model Hamiltonian by 
matching wavefunctions at the $z=\pm T/2$ boundaries~\cite{SM}.

The Dirac equation solutions include a set of non-chiral eigenfunctions whose role we focus on in this paper, 
and a chiral eigenfunction with velocity $v_\mathrm{D} =\sqrt{v_{\mathrm{D}z} v_{\mathrm{D}y}}$, energy $E(k)=\hbar v_\mathrm{D}k$,
and a wavefunction that is constant inside the side wall and decays exponentially on the top and bottom surfaces 
[See Fig.~\ref{fig4}(b)].  The non-chiral eigenvalues are conveniently expressed in dimensionless units 
related to the sidewall's size-quantization energy scale:
$\varepsilon=E T/\pi \hbar v_{\mathrm{D}z}$, $\mu=m_0 T/\pi \hbar v_{\mathrm{D}z}$ and $\chi=k \sqrt{v_{\mathrm{D}y}/v_{\mathrm{D}z}} T/\pi$. 
Because the non-chiral band energies are even functions of $\chi$, 
non-chiral states always appear in equal-energy,
opposite-velocity, opposite-wavevector pairs. 
The number  $N_{\rm NC}$
of non-chiral one-dimensional subbands that are occupied at energy $\varepsilon$ 
decreases with dimensionless mass $\mu$, as illustrated in Fig.~\ref{fig4}(c) where the 
energies of non-chiral band minima, located at $\chi=0$, are plotted as a function of $\mu$. 
The wavefunctions of the non-chiral states  
have nodes along the $z$-direction, with the number of nodes increasing with the energy of the 
state [Fig.~\ref{fig4}(d)].  
For thick films, the number of sidewall channels at energies inside the surface-state gap is 
$N_{\rm NC} \sim \mu = m_0 T/\pi \hbar v_{\mathrm{D}z} \sim  (m_0 [{\rm meV}] \, T[{\rm nm})]/ (200 \, v_{\mathrm{D}z} [{\rm 10^{5} m/s}])$. 
It follows that for $\mu \lesssim 1$, non-chiral states are absent across most of the surface state gap.  
Non-chiral channels are present across a larger fraction of the gap for thicker films, larger gaps, and 
smaller vertical Dirac velocities.  Below we confirm these predictions of the simplified toy model, 
and obtain a numerical estimate for $v_{\mathrm{D}z}$ by performing microscopic tight-binding-model calculations.  

\noindent
\textit{Microscopic Sidewall State Theory}---
In order to address transport in the QAHE regime, it 
is necessary to study the sidewall electronic structure microscopically~\cite{ChenAPL2015,WangPhysRevLett.111.086803}.
This will allow us: i) to determine the velocity parameter $v_\mathrm{Dz}$ that along with the film thickness 
sets the sidewall finite-size quantization
energy scale; ii) to examine the position of the Dirac point relative to the bulk conduction and valence bands,  
and iii) to identify and shed light on relevant features (addressed below) that are not 
captured by simple continuum models. 
We focus on Bi$_2$Se$_3$ family~\cite{Hsieh2009} topological insulators.
Electron states in this crystal can be  
described by a \textit{sp}$^3$ tight-binding model 
with parameters obtained by fitting 
to \textit{ab initio} 
calculations~\cite{Kobayashi,NJP}. 
In order to model homogeneous perpendicular magnetization, we introduce an exchange field
$B_{\rm ex}$ expressed in energy units and oriented perpendicular to the ($111$) surface.
We comment below on the relationship between $B_{\rm ex}$ and the mass parameter $m_0$ of the sidewall state toy model.   

\begin{figure}[ht!]
\centering\includegraphics[width=0.98\linewidth,clip=true]{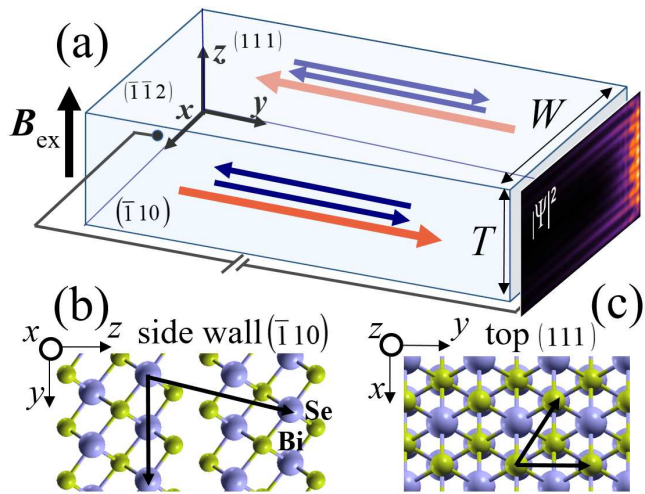}
\caption{(Color online) (a) Schematic of a Bi$_2$Se$_3$ nanoribbon with finite thickness $T$ and 
width $W$.  The thick red arrows represent currents carried 
by ballistic chiral edge states, while the thin blue arrows represent currents carried 
by ballistic non-chiral edge state channels.  The sidewall localization property of 
chiral states is reflected by the illustrated $|\Psi|^2$ probability density distribution
across the ribbon cross section.  Top views of the ($\bar{1}10$) sidewall (b) and
the ($111$) top and bottom surface layers (c), 
with black arrows for 2D crystal unit vectors.} 
\label{fig1}
\end{figure} 

\begin{figure}[ht!]
\centering
\includegraphics[width=0.98\linewidth,clip=true]{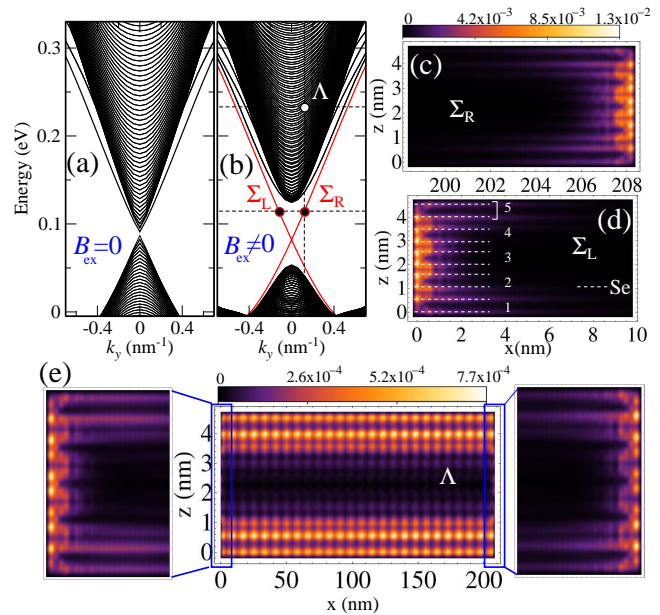}
\caption{(Color online) Bandstructure of a Bi$_2$Se$_3$ nanoribbon with $T=5$QL and $W=208$~nm for 
$B_{\rm ex}=0$ (a) and $B_{\rm ex}=0.16$~eV (b). 
Chiral edge states are shown in red in panel (b). Spatial distribution of the wavefunction 
across the ribbon cross-section for a right-goer state $\Sigma_R$ (c) and a left-goer state $\Sigma_L$ (d) with energy
$\approx 0.11$~eV [dashed line in (b)], and a non-chiral state $\Lambda$ (e) with energy $\approx 0.23$~eV 
[white circle in (b)]. 
The dashed lines in (d) mark the positions of the outermost Se layers in each QL. 
In (c) and (d)  only the ribbon edges are shown.  For this ribbon geometry there are no non-chiral channels 
that are inside the surface state gap and localized on the sidewalls.} 
\label{fig2}
\end{figure}

To extract the facet-dependent surface state Dirac velocities we first consider the infinite cross-sectional-area 
thin-film geometry.  For the Se ($111$) surface-layer facet,
we find that the Dirac cone is isotropic with velocity
$v^0_\mathrm{D}\approx 5.0\times 10^5$~m/s.  For the ($\bar{1}10$) sidewall facet,
we find that the Dirac cone is strongly anisotropic with 
$v_{\mathrm{D}y}\approx 4.8\times 10^5$~m/s and $v_{\mathrm{D}z}\approx 2.3\times 10^5$~m/s.
We then turn to the ribbon geometry [Fig.~\ref{fig1}(a)] in order to identify the 
side-wall states active in quantum Hall transport experiments.  
The ribbon is infinite in the $y$-direction, the direction of longitudinal transport, 
has a thickness $T$ in the $z$-direction approximately equal to 1 nm per quintile layer (QL), 
and a finite width $W$ in the $x$-direction. 
The mixed Bi/Se sidewall ($\bar{1}10$) surfaces are illustrated in Fig.~\ref{fig1}(b), 
and the Se ($111$) top and bottom surfaces in Fig.~\ref{fig1}(c)~\cite{LeePNAS2015}.
Results for ribbons with $T=5$QL and $W=208 {\rm nm}$ are presented in Fig.~\ref{fig2}.

At $B_{\rm ex}=0$ the low-energy states consist of discrete
quasi-1D channels that are
separated in energy by $\sim \hbar \pi v_\mathrm{D}/(T+W)$~\cite{BardasonPhysRevLett.105.156803}
as illustrated in Fig.~\ref{fig2}(a). 
Wavefunctions at energies within the bulk 
gap, roughly between 0 and 0.4~eV for a 5QL film, 
are distributed over all four facets of the ribbon at $k=0$, 
but because of the Dirac velocity mismatch
tend to localize either on sidewall or on surface facets at $k \ne 0$. 
At exchange field $B_{\rm ex}=0.16$eV a gap opens and is bridged by a pair of chiral edge 
states [Fig.~\ref{fig2}(b)].    
The size of the gap is smaller than the exchange coupling energy 
because, in contrast to the toy model, the quasiparticles have mixed spin character even at $k=0$ and $g$-factors 
that are smaller than $2$.  The gap in Fig.~\ref{fig2}(b) is $\sim 0.07 eV$ and can be identified with the 
gap $2 m_0$ in the surface state toy model.  It follows that $B_{\rm ex} = 0.16$eV corresponds to 
$m_0 \sim 35 {\rm meV}$, in agreement with typical experimental estimates. At a given energy, 
the negative-velocity ($\Sigma_\mathrm{L}$) and positive-velocity ($\Sigma_\mathrm{R}$) 
states (inside the surface state gap) are localized on the opposite side walls [see Fig.~\ref{fig2}(c) 
and (d)]. The value of the chiral-state velocity extracted from the 
microscopic calculation is $\approx 4.1 \times 10^{5}$ m/s, which
is consistent with the estimate provided by the toy-model expression 
$v_\mathrm{D} =\sqrt{v_{\mathrm{D}z} v_{\mathrm{D}y}} \approx  3.3 \times 10^{5}$ m/s  
(evaluated for $v_{\mathrm{D}z}$ 
and $v_{\mathrm{D}y}$ quoted above).   
Both states are spin-polarized in the direction of the exchange field. 

The chiral edge state wavefunctions are evenly distributed across the 
sidewalls apart from variations that mirror the chemical structure of the QLs
and place larger weight on the Se layers. 
Chiral states are more strongly localized on the sidewalls at larger $k$. 
Non-chiral channels appear in this calculation only at energies outside of the 
surface state gap.  A typical non-chiral state, illustrated in Fig.~\ref{fig2}(e) ($\Lambda$),
has weight on both sidewalls and on the top
and bottom surfaces.  

The absence of side-wall localized non-chiral transport channels in these calculations 
can be understood by comparing with the side-wall toy model introduced in 
the previous section and using the microscopically
calculated value for $v_{\mathrm{D}z}$ to evaluate the dimensionless 
gap parameter.  We find that for the thickness and exchange interaction strength
of this representative microscopic calculation $\mu \sim 0.3$, consistent with the  
$N_{\mathrm{NC}}=0$ electronic structure of Fig.~\ref{fig2}.(b)  Non-chiral channels 
appear at energies inside the surface 
state gap only for thicker films or stronger exchange splitting.  

\begin{figure}[ht!]
\centering
\includegraphics[width=0.98\linewidth,clip=true]{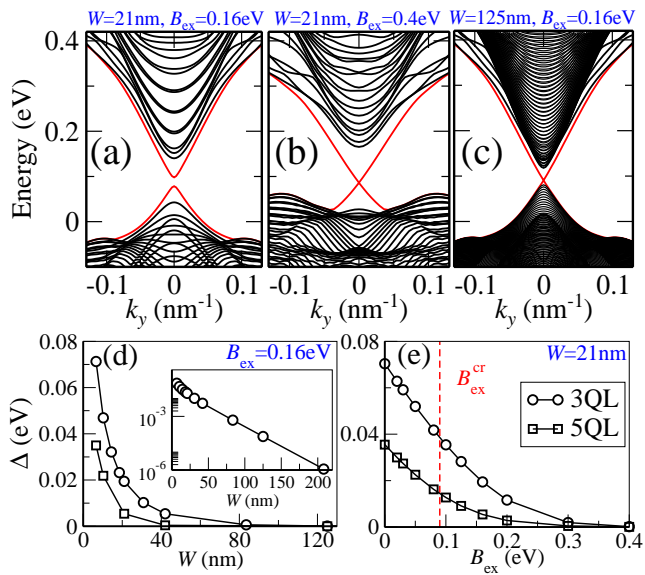}
\caption{(Color online) Electronic structure of a Bi$_2$Se$_3$ nanoribbon with $T=3$QL for $W=21$~nm and $B_{\rm ex}=0.16$~eV (a), 
$W=21$~nm and $B_{\rm ex}=0.4$~eV (b), and $W=125$~nm and $B_{\rm ex}=0.16$~eV (c). Chiral edge states are shown in red. 
(d) Chiral state avoided crossing gap for $T=3$QL (circles) and $T=5$QL (squares) as a function of $W$ for $B_{\rm ex}$=0.16~eV. 
The inset shows the logarithm of the gap for 3QL films over a larger range of $W$'s. 
(e) Energy gap as a function of $B_{\rm ex}$ for a fixed $W=21$~nm  
in the same two cases. The vertical dashed line marks the critical exchange field ($B_{\rm ex}^{\rm cr}\approx 0.09$~eV) for 3QL.} 
\label{fig3}
\end{figure}  

\noindent
\textit{Quantum Hall Transition in Very Thin Films}--- 
Our sidewall toy model does not account for the hybridization between top and bottom 
surfaces which, in very thin films, can control a transition between quantum Hall and topologically trivial states.~\cite{Yu02072010} 
The hybridization scale is negligible compared to typical exchange energy scales in the 5QL films 
discussed above, but not in the 3QL films whose properties are summarized in Fig.~\ref{fig3}.
Hybridization plays an essential role in 3QL films by opening a sizable surface state gap at $B_{\rm ex} = 0$. 
This time-reversed ground state of the 3QL film is a two-dimensional topological insulator and supports helical edge states. 
The gap decreases in size with increasing $B_{\rm ex}$ and 
vanishes at $B_{\rm ex} = B_{\rm ex}^{\rm cr}$.  For $B_{\rm ex} > B_{\rm ex}^{\rm cr}$ the order of 
the lowest two-dimensional subbands is reversed, causing a transition to the QAHE phase~\cite{SM}, 
 and the gap size then increases with $B_{\rm ex}$.  
We find that $B_{\rm ex}^{\rm cr}\approx 0.09$~eV for 3QL films and that  
$B_{\rm ex}^{\rm cr}\approx 10^{-2}$~meV for 5QL films.
Although remnants of the $B_{\rm ex}=0$ helical edge states can complicate 
edge electronic structure when  $B_{\rm ex} \lesssim B_{\rm ex}^{\rm cr}$, 
our microscopic calculations demonstrate that no trace is present for
$B_{\rm ex} \gg B_{\rm ex}^{\rm cr}$ where only the chiral edge modes survive.  

For finite width ($W$) ribbons there is a finite gap in the electronic structure 
for $B_{\rm ex} > B_{\rm ex}^{\rm cr}$ because of the avoided crossing 
between edge states localized on opposite side walls.   
The ribbon gap decreases in size both with increasing $B_{\rm ex}$ [Fig.~\ref{fig3}(b)]
and increasing $W$ [Fig.~\ref{fig3}(c)]. 
For a fixed exchange field $B_{\rm ex}>B_{\rm ex}^{\rm cr}$ 
the energy gap decreases exponentially with $W$ [Fig.~\ref{fig3}(d)], 
whereas for $B_{\rm ex} < B_{\rm ex}^{\rm cr}$, the gap approaches a finite value as 
$W \to \infty$.  
By fitting the $W$ dependence of the gap to an exponential 
decay law 
we estimate that the localization length 
of the chiral edge state at $ B_{\rm ex} = 0.16 eV$
is $\lambda\approx 18.6$~nm for 3QLs and $\lambda\approx 8.2$~nm for 5QLs.
Since typical experimental samples used in quantum anomalous Hall studies have 
widths of hundreds of $\mu$m, direct coupling between opposite edges 
is negligible in the absence of disorder.  

\noindent
\textit{Quantum Anomalous Hall Effect}---
Experimental QAHE measurements have so far been performed mainly on films with 
thicknesses in the range between 5 and 10 QL.  Because the vertical sidewall Dirac velocity, which characterizes 
a direction in which electrons hop between Bi and chalcogen layers, is only a few times smaller than Dirac velocities 
in directions along Bi layers, we conclude that the 10 QL layer maximum thickness is not sufficient to support non-chiral edge modes.   
At the same time, hybridization between top and bottom surfaces at the minimum 5QL thickness is very much weaker 
than typical exchange fields.  For this reason, we conclude that the sidewalls of the samples that are typically 
studied do not support either helical edge states that are a remnant of $B_{\rm ex}=0$ two-dimensional 
topological insulator states, or the non-chiral side wall states~\cite{WangPhysRevLett.111.086803} that are expected in thicker films. 
The case in which the surface state Dirac point is buried inside the valence band of the host topological insulator 
might provide an exception to these conclusions, but is not in any case expected to be ideal for the realization of the 
QAHE.

Because the sidewall spectrum of the QAHE samples consists of a single 
chiral channel, it is not possible to explain the commonly observed finite longitudinal resistances 
by assuming a failure to establish local equilibrium on a multi-channel edge.  A more likely explanation, 
in our view, is that potential disorder causes the local Fermi level to sweep across the surface state gap.  
The relatively high-velocity one-dimensional chiral sidewall states have negligible density of states. Therefore they
can do little on their own to screen inevitable spatial variations in external electric fields that 
induce relative shifts in the Dirac cones of top and bottom surfaces, or external potentials that induce common shifts 
in the Dirac cones of the two surfaces.  Fluctuations that bring the surface states to the Fermi level, provide a 
mechanism for two-dimensional dissipative transport in some parts of the system. Because the surface states
themselves have a large~\cite{Sinitsyn2007, Sinitsyn2008}, but unquantized Hall conductivity in addition to a  
finite longitudinal conductivity, surface conduction will tend to lead more to a finite longitudinal 
conductivity than to 
a correction to the Hall conductivity.

The quantum Hall effect may be more robust against disorder in thicker films, if they can be grown while maintaining similar 
sample quality.  Because the edge carries current in equilibrium, the presence of many non-chiral channels does not lead either 
to inaccuracies in Hall quantization or to longitudinal resistance.  Instead a larger number of channels at the edge 
increases the degree to which disorder is screened and helps broaden the gate voltage range over which 
nearly pure side wall transport can be established.  

This work was supported by the Faculty of Technology at Linnaeus University and by the
Swedish Research Council under Grant Number: 621-2014-4785. 
AHM was supported by the Welch Foundation under grant F-1473 and by
SHINES, an Energy Frontier Research Center funded by the U.S. Department of Energy (DoE), Office of Science, Basic Energy Science (BES) under award  DE-SC0012670.  AHM acknowledges valuable interactions
with Yabin Fan, Massoud Masir, Pramay Upadhyaya, Kang Wang, Fengcheng Wu and Fei Xue.
Computational resources have been provided by the Lunarc center for scientific and technical computing at
Lund University.

\bibliography{pertsova}
\end{document}